\documentclass[]{spie}  

\usepackage{amsmath,amsfonts,amssymb}
\usepackage{graphicx}
\usepackage[colorlinks=true, allcolors=blue]{hyperref}
\usepackage{wrapfig}

\usepackage{placeins}
\usepackage{caption}
\usepackage{subcaption}
\usepackage{graphbox}
\usepackage[export]{adjustbox}
\usepackage{gensymb}
\usepackage{amsmath}
\usepackage{wrapfig}

\title{Combining Spectral CT Acquisition Methods \\for High-Sensitivity Material Decomposition }
\author[]{\normalsize\vspace{-3mm} Matthew Tivnan$^1$, Wenying Wang$^1$, Grace J. Gang$^1$, Eleni Liapi$^2$, Peter No\"{e}l$^3$, J. Webster Stayman\vspace{-3mm}}
\affil[1]{\normalsize\em Department of Biomedical Engineering, Johns Hopkins University, Baltimore, MD, 21205}
\affil[2]{\normalsize\em Department of Radiology, Johns Hopkins University, Baltimore, MD, 21205}
\affil[3]{\normalsize\em Department of Radiology, Perelman School of Medicine, University of Pennsylvania, Philadelphia, PA, 19104}

\begin{document} 
\maketitle
\vspace{-6mm}
\begin{abstract}
\vspace{-2mm}
Quantitative estimation of contrast agent concentration is made possible by spectral CT and material decomposition. There are several approaches to modulate the sensitivity of the imaging system to obtain the different spectral channels required for decomposition. 
Spectral CT technologies that enable this varied sensitivity include source kV-switching, dual-layer detectors, and source-side filtering (e.g., tiled spatial-spectral filters). In this work, we use an advanced physical model to simulate these three spectral CT strategies as well as hybrid acquisitions using all combinations of two or three strategies. We apply model-based material decomposition to a water-iodine phantom with iodine concentrations from 0.1 to 5.0 mg/mL. We present bias-noise plots for the different combinations of spectral techniques and show that combined approaches permit diversity in spectral sensitivity and improve low concentration imaging performance relative to the those strategies applied individually. Better ability to estimate low concentrations of contrast agent has the potential to reduce risks associated with contrast administration (by lowering dosage) or to extend imaging applications into targets with much lower uptake.
\end{abstract}


\vspace{-5mm}
\section{INTRODUCTION}
\label{sec:intro}  
\vspace{-1mm}

Contrast-enhanced computed tomography is an important clinical tool for imaging the three-dimensional structure and function of tissues in the human body. Applications include angiography, perfusion studies, and imaging of contrast-enhanced tissues. In many cases, small changes in contrast agent (e.g. iodine) have important clinical implications. Hepatocellular carcinoma, for example, is known to uptake slightly higher concentrations during the arterial phase relative to healthy liver tissue and it is also known to return to a low-concentration state slightly faster than healthy tissue during the venous phase \cite{choi2014ct}. Sensitivity to low concentrations of iodinated contrast agents in the presence of noise will allow the next generation of CT systems to convey more functional information, e.g. regions of lower uptake, or to conduct similar studies with less contrast injection.

Spectral CT is finding increasing application in contrast-enhanced CT due to its ability to discriminate between contrast and human tissues and its ability to provide quantitative estimates of contrast concentration. Spectral CT uses projection measurements with varied spectral sensitivities to enable material-decomposed image volumes (material density maps). For example, since iodine and soft-tissue have differently-shaped mass attenuation spectra as a function of energy, spectral CT systems can take advantage of this additional information to quantify concentration.

\vspace{-3mm}

\begin{figure}[h!]
    \centering
    \includegraphics[height=0.28\textwidth,trim = {4mm 2mm 2mm 0},clip]{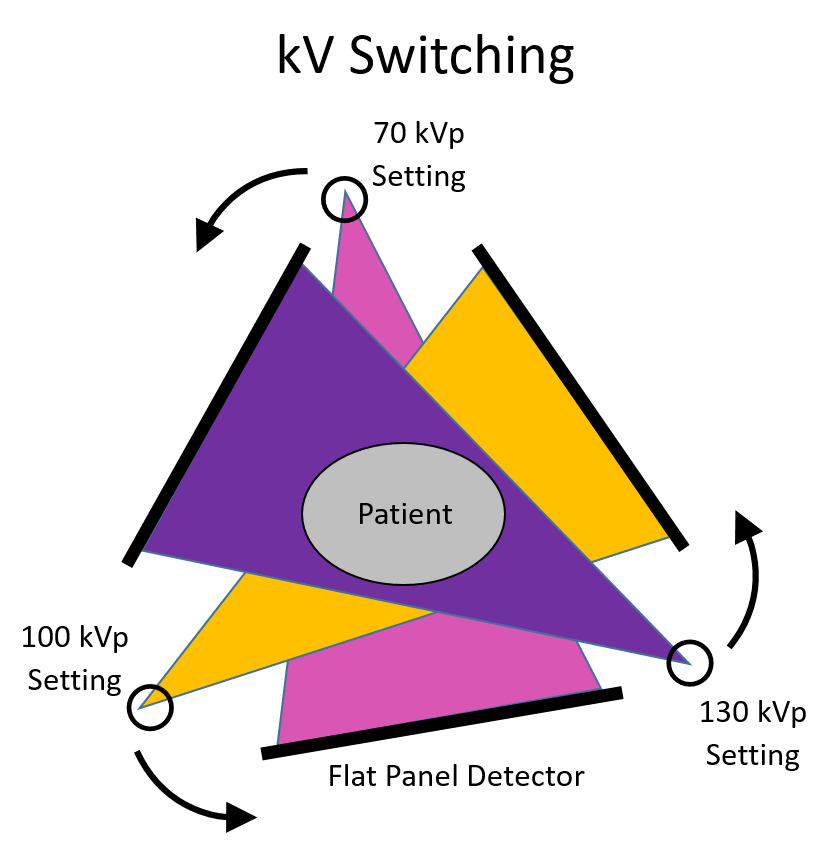}
    \includegraphics[height=0.28\textwidth,trim = {4mm -10mm 4mm 0},clip]{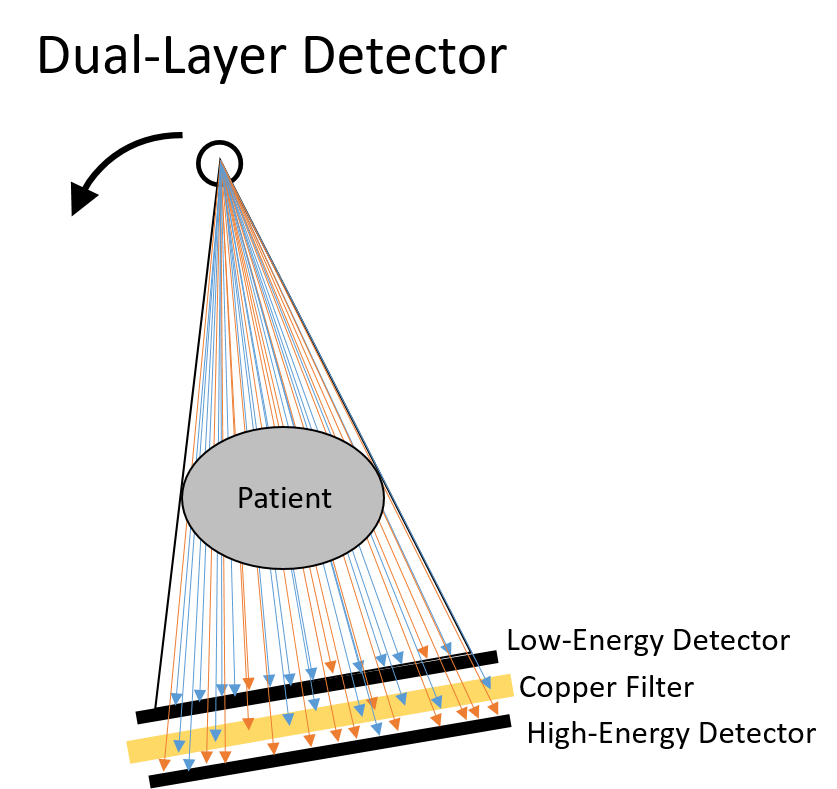}
    \includegraphics[height=0.28\textwidth,trim = {4mm -2mm 2mm 0},clip]{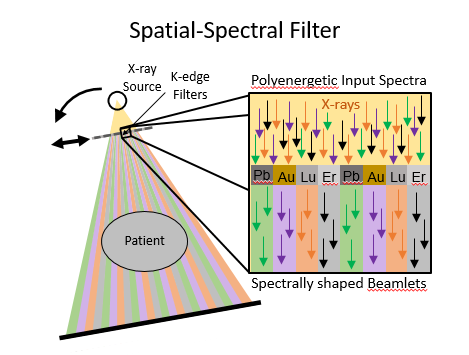}
    \caption{Illustrations for kV Switching (left) Dual-Layer Detectors (center) and Spatial-Spectral Filters (right).}
    \label{fig:methods_cartoon}
\end{figure}

\vspace{-3mm}

There are various strategies that enable Spectral CT. For example, photon counting detectors are capable of energy discrimination and can effectively break projection data into different spectral channels (i.e., energy bins). 
Source-switching schemes are also used, wherein the voltage across the x-ray tube alternates between two or more kVp settings \cite{zou2008analysis}.  In this fashion kV-switching (KVS) provides angular projections with a diversity of incident spectra. Split-filters, which divide the source beam into a low-energy and high-energy half-cone beams, yield two spectral channels for dual-energy CT \cite{primak2009improved}. More recently, a generalization of the split filter idea called a spatial-spectral filter (SSF) has recently been proposed \cite{tivnan2019physical}. The SSF consists of a tiled array of k-edge filters which divide the beam into a number of filtered beamlets to provide a diversity of energy channels. See Figure \ref{fig:methods_cartoon} for illustrations of these spectral CT methods.

We hypothesize that combinations of spectral data acquisition approaches have the potential to increase spectral CT sensitivity to low concentrations of contrast agent. That is, with a greater diversity of spectral channels, one can improve sensitivity over the use of any single individual approach.  Many approaches may be combined (e.g. KVS and SSF) into a hybrid acquisitions for potential improvements in performance. Historically, these combinations may have been avoided since it has the potential to greatly complicate data processing. For example, traditional image- or projection-domain processing can be challenged by beam-hardening effects or projection sampling requirements - e.g., combining KVS and SSF makes channelized data with unique spectral sensitivity increasingly sparse and difficult to reconstruct individually. However, recent advances in direct model-based material decomposition (MBMD) \cite{tilley2018model} enable such hybrid acquisitions since all spectral effects for each measurement can be modeled and there is no intermediate processing - material density maps are formed directly from the varied spectral projection data using model-based iterative approaches. 

In this work we present a study where we employ a detailed physical model of various spectral CT strategies appropriate for combination on a CT system with an energy-integrating detector including kV-switching, dual-layer detectors (DLD) \cite{lu2019dual}, and spatial-spectral filters. We apply MBMD to simulated data using a digital water-iodine phantom, and evaluate performance for individual spectral approaches as well as two- and three-strategy hybrid acquisitions to analyze the impact of spectral diversity on material decomposition performance for contrast-enhanced spectral CT.

\FloatBarrier
\vspace{-1em}
\section{METHODS}

\subsection{General Spectral CT Forward Model and Model-Based Material Decomposition}

A general forward model for spectral CT with varied spectral measurements and energy-integrating detectors is


\setlength{\abovedisplayskip}{0pt}
\setlength{\belowdisplayskip}{3pt}
\setlength{\abovedisplayshortskip}{0pt}
\setlength{\belowdisplayshortskip}{3pt}
\vspace{-3mm}
\begin{equation}
\vspace{-3mm}
	\bar{y_i} = g_i \int_E S_i(E) \exp{\Big(-\sum_j a_{ij}  \mu_j(E) \Big)} dE \enspace, \quad \quad \quad \mu_j(E) = \sum_k  {\rho}_{j,k} q_k(E) 
\end{equation}

\noindent where $\bar{y_i}$ is the mean value of the $i^{th}$ measurement, $g_i$ is the corresponding gain, $S_i(E)$ is the measurement-dependent system spectral sensitivity (including detector energy sensitivity, filtration, x-ray source spectrum, etc.), $a_{ij}$ are projection contributions of the $j^{th}$ voxel to the $i^{th}$ measurement, and $\mu_j(E)$ is the total mass attenuation spectrum of the $j^{th}$  voxel. The latter spectrum is modeled as a weighted sum over $k$ basis materials with material mass attenuation basis functions $q_k(E)$ weighted by material densities, $\rho_{j,k}$ for each voxel. In practice, we discretize energy spectra as follows

\begin{wrapfigure}[11]{r}{0.55\textwidth}
    \centering
    \vspace{-1mm}
    \includegraphics[width=0.27\textwidth,trim = {5mm 2mm 11mm 0},clip]{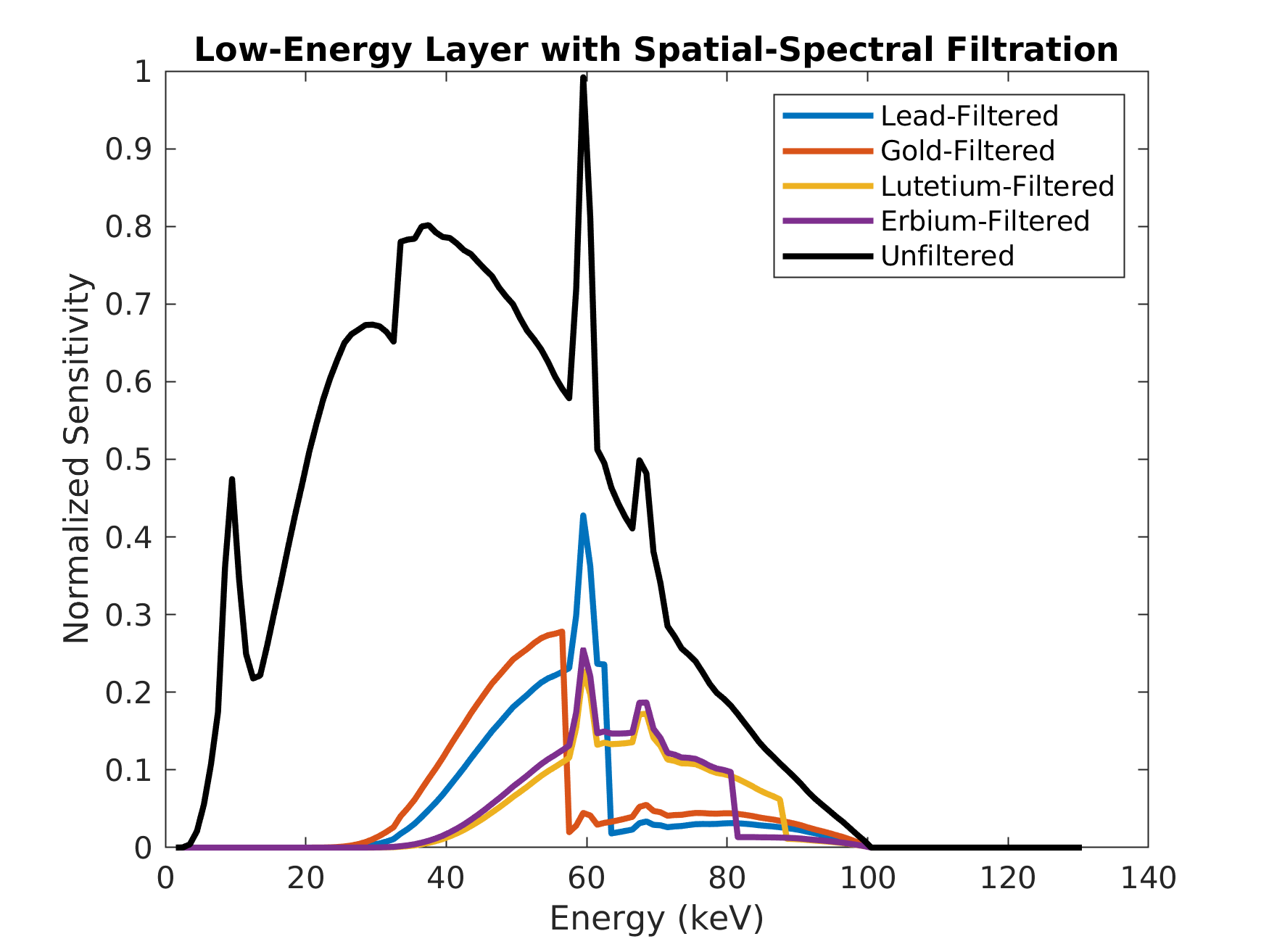}
    \includegraphics[width=0.27\textwidth,trim = {5mm 2mm 11mm 0},clip]{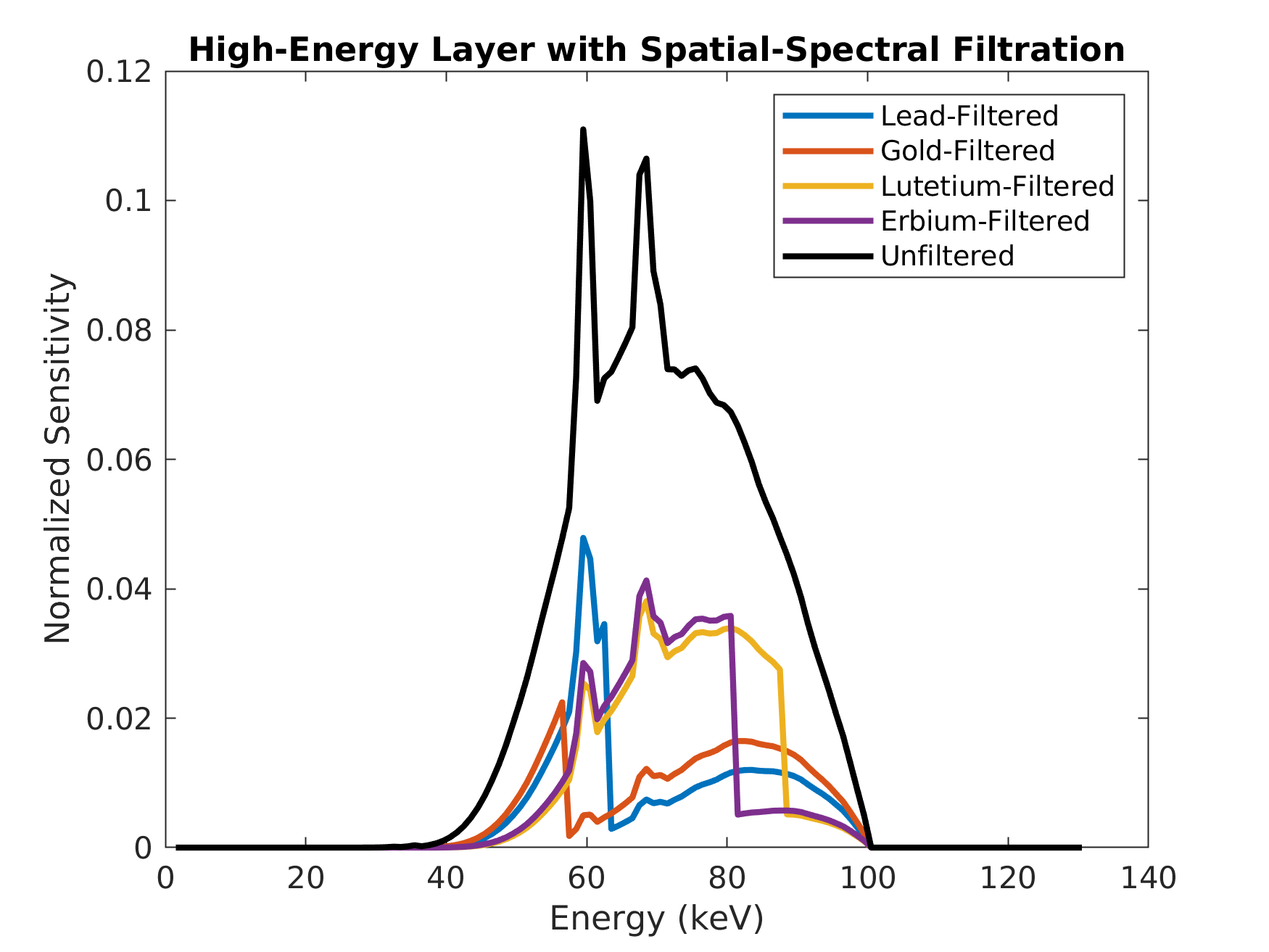}
    \caption{DLD and SSF hybrid spectral sensitivities.}
    \label{fig:DLD_SSF_spectra}
\end{wrapfigure}

\vspace{-3mm}
\begin{equation}
\vspace{-2mm}
	\bar{y_i} = g_i \sum_w s_{iw} \exp{\Big(-\sum_k q_{kw} \sum_j a_{ij}  {\rho}_{j,k} \Big)}  \quad \text{or, compactly, as} \quad \bar{\mathbf{y}} = \mathbf{G} \mathbf{S} \exp{\Big(- \mathbf{Q} \mathbf{A} \boldsymbol{\rho} \Big)}
\end{equation}
This model is extremely general. For example, for spatial-spectral filters we can use a high-resolution voxelized model of the filter tile array and perform a forward projection to establish $S_i(E)$ as shown in Figure \ref{fig:DLD_SSF_spectra}. There is also no restriction to have a small number of discrete spectral channels which allows us to model many combinations of spectral effects as well as calibrate physical non-idealities (e.g. spectral mixing at filter boundaries).


%

To conduct MBMD, we seek to minimize the following penalized weighted least-squared objective function:

\vspace{-5mm}
\begin{equation}
\vspace{-3mm}
	\Phi = \sum_i w_i (\bar{y_i} - y_i)^2 + \beta_0 R\{\boldsymbol{\rho}\}
\end{equation}

\noindent where $w_i$ are the statistical weights, for which we substitute $y_i$ (approximate Poisson noise model), and $R\{\cdot{}\}$ is a regularization term. In this work we adopt a quadratic penalty with both  in-basis and cross-basis penalties\cite{wang2019local} to minimize cross-talk between material basis. Moreover, we seek matched spatial resolution between basis materials, and we control the overall penalty weight via $\beta_0$. We derive the gradient, $\nabla \Phi$, as well as the Hessian, $\mathbf{H}$, of $\Phi$ with respect to $\boldsymbol{\rho}$ and we apply the following Newton's Method iterative update scheme:

\vspace{-5mm}
\begin{equation}
\vspace{-3mm}
	\boldsymbol{\rho}^{(n+1)} = \boldsymbol{\rho}^{(n)} - \alpha^{(n)}{\mathbf{H}^{(n)}}^{-1}\nabla\Phi^{(n)}
\end{equation}

\noindent where $\alpha^{(n)}$ is a step size chosen in such a way as to satisfy Wolfe's strong conditions on convergence. The cross-voxel second derivatives contained in $\mathbf{H}$ are approximated to be zero. That is, $\frac{\partial}{\partial \rho_{j_1 k}}\frac{\partial}{\partial \rho_{j_2 k}} \Phi = 0$ for $j_1 \neq j_2$. However, the cross material second derivatives $\frac{\partial}{\partial \rho_{j k_1}}\frac{\partial}{\partial \rho_{j k_2}} \Phi $ are included to accommodate, in the search direction, the strong anticorrelation between basis materials.

\vspace{-3mm}

\subsection{Simulation Study on Hybrid Acquisitions}
\vspace{-1mm}

To characterize the relative performance of KVS, DLD, and SSF technologies as well as every combination of two or three (e.g. KVS-DLD, or KVS-DLD-SSF), we conducted spectral CT simulations and material decomposition experiments with a 1.0~mm voxelized water/iodine phantom containing a 100.0~mm diameter water tank and six 10.0~mm diameter iodine contrast inserts of concentration 0.10, 0.25, .50, 1.00, 2.50, and 5.00~mg/mL. We modeled a CT system with a source-to-detector distance of 1200~mm, a source-to-axis distance of 600~mm, and 360 views per rotation. For the standard detector, we modeled a flat panel with a 600~$\mu$m CsI scintillator and a 1.112~mm pixel pitch. The source spectra were modeled using the TASMICS dataset via the software package SPEKTR 3.0 \cite{punnoose2016spektr}.  

For the three spectral CT strategies, we chose a KVS scheme which involved switching between 70, 100, and 130~kVp settings in that order view-by-view such that each spectral channel is made up of 120 evenly spaced full views. For the DLD system, we replaced the default flat-panel detector model with a Varex prototype dual-layer flat-panel detector which consists of an initial low-energy detector layer with a 175~$\mu$m CsI scintillator, followed by a 1.0~mm copper filter for spectral shaping, and a second layer with a 550~$\mu$m CsI scintillator for the high-energy channel. For the SSF system we model a source-to-filter distance of 100.0~mm and 3.0~mm filter tiles with a linear translation speed of 0.4~mm per view. The filter tile materials are lead, gold, lutetium, and erbium (chosen for their k-edges). All filters were 0.25~mm thick with the exception of the gold filter tile which was 0.15~mm thick.

For fair comparisons between each individual and combined spectral CT system, we modified tube current so as to normalize the deposited dose, which we computed using a uniform 160.0~mm water cylinder model of the patient. The maximum detector efficiency in the study occurs with the SSF system acting alone and is adjusted to a level of $10^6$ total photon counts per pixel per view. We used a Poisson additive model to simulate noise.

\begin{wrapfigure}[14]{r}{0.7\textwidth}
    \centering
    \vspace{-4mm}
    \includegraphics[height=0.28\textwidth,trim = {24mm 10mm 10mm 6mm},clip]{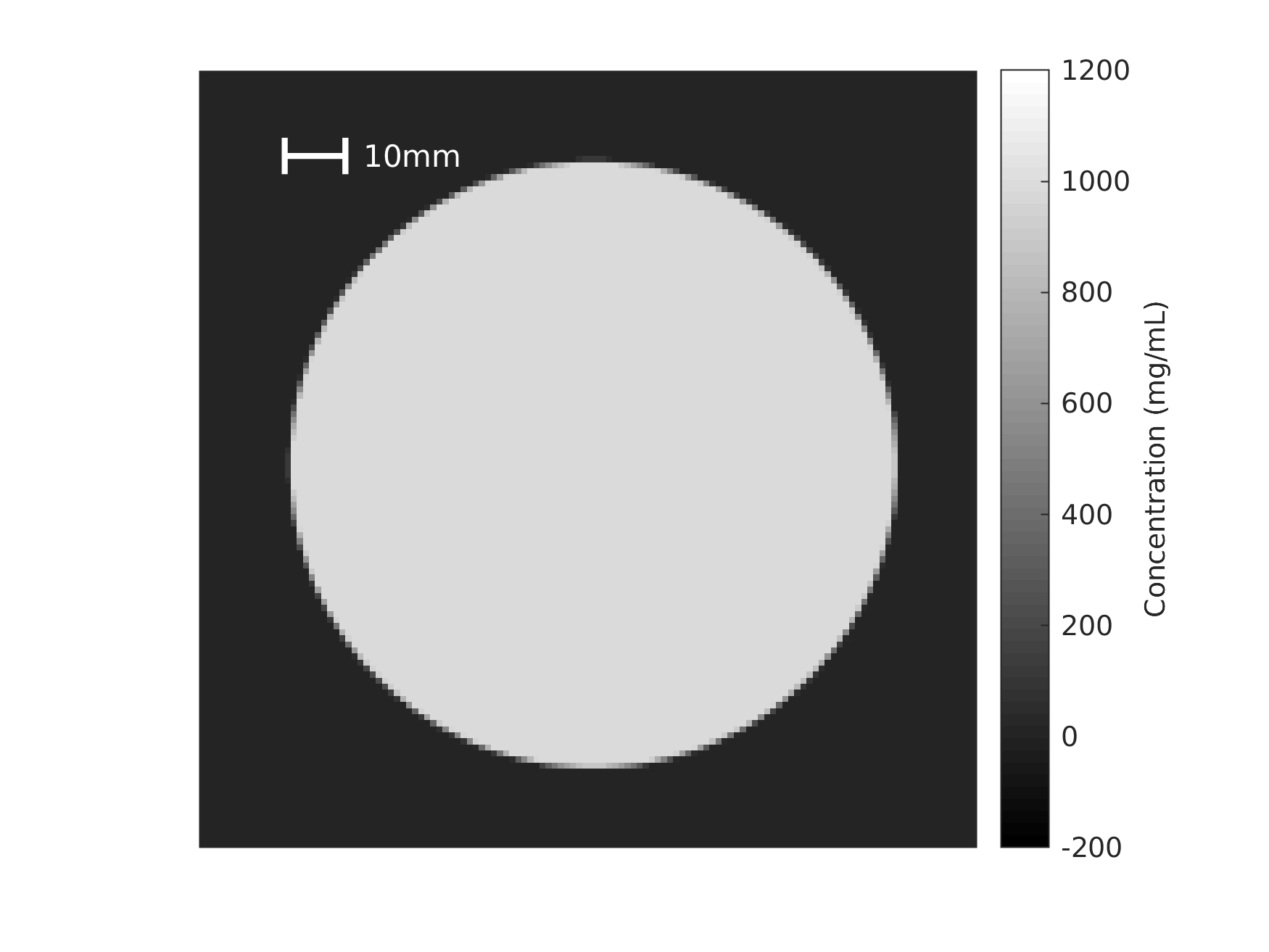}
    \includegraphics[height=0.28\textwidth,trim = {24mm 10mm 16mm 6mm},clip]{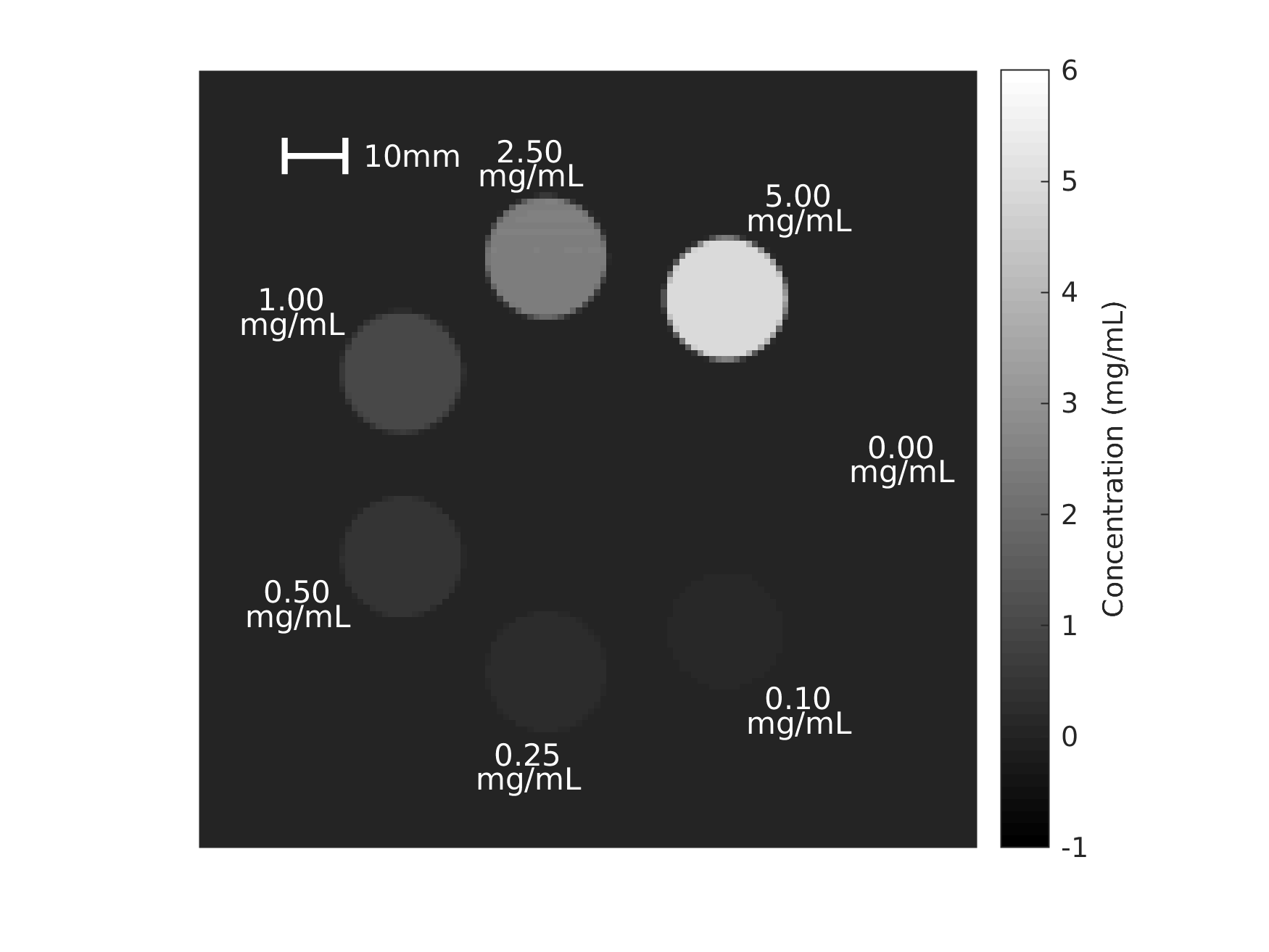}
    \caption{Ground truth numerical phantom. Water (left) Iodine (right) }
    \label{fig:phantom_GT}
\end{wrapfigure}

For each acquisition method, we ran 300 iterations of the Newton's Method material decomposition algorithm. A range of penalty weights, $\beta_0$, was applied to investigate the bias-noise trade-off for each strategy. That is, for small penalty weights one should see high variance and low bias, and for large penalty weights, low variance and high bias (due to blurring of edges). Noise standard deviation was computed in an 8.0~mm diameter ROI internal to the 10.0~mm diameter insert containing a uniform 5.00~mg/mL of iodine as shown in Figure \ref{fig:phantom_GT}. To compute bias, we run an additional noiseless material decomposition ans we compute RMSE over a 14.0~mm diameter ROI centered on the 5.00~mg/mL insert. This ROI metric is intended to capture the bias due to the blurring of edges and any other errors excluding noise.



\begin{figure}[h!]
    \centering
    \includegraphics[height=0.34\textwidth,trim = {24mm 2mm 16mm 0},clip]{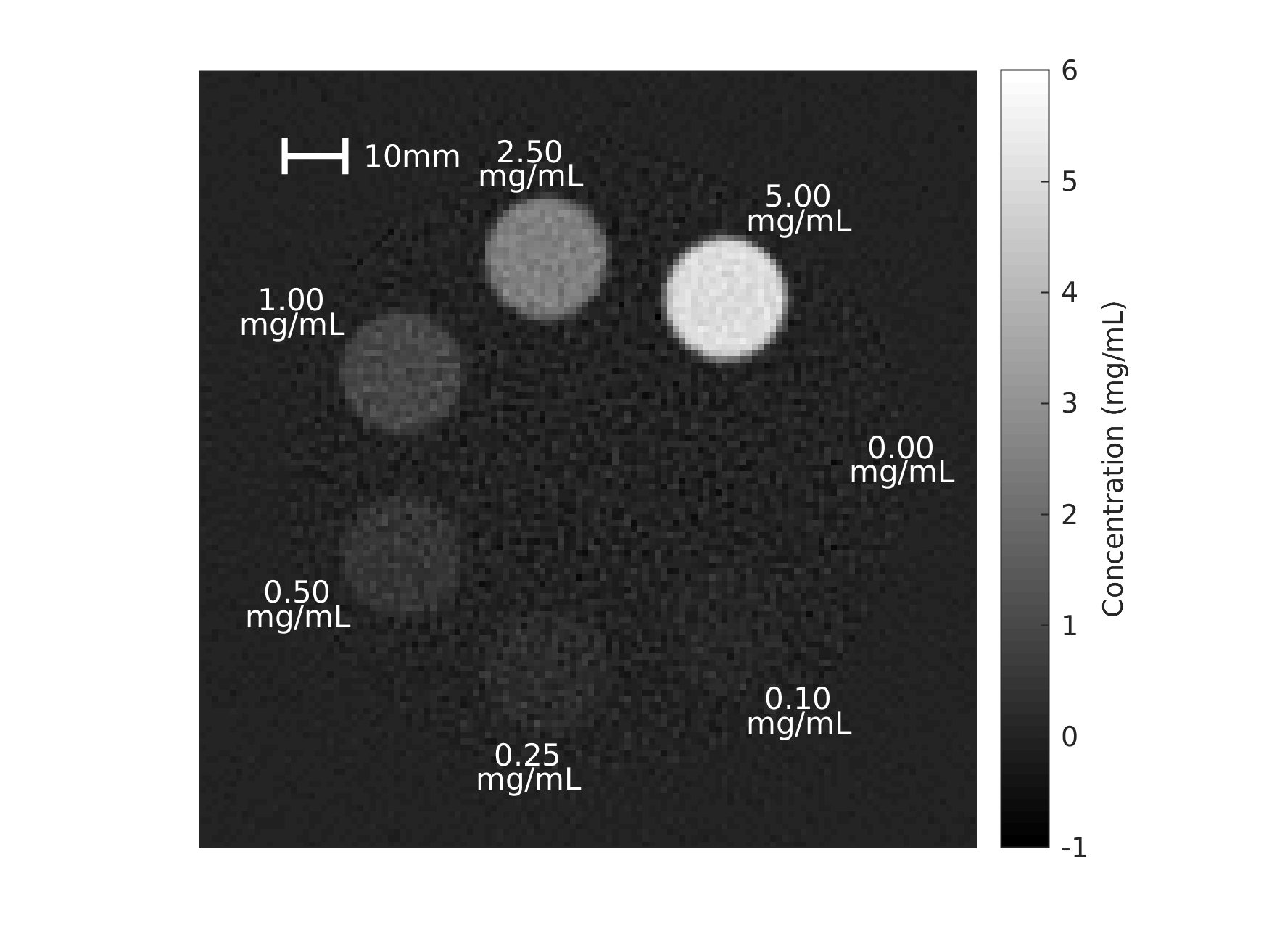}
    \includegraphics[height=0.34\textwidth,trim = {24mm 2mm 32mm 0},clip]{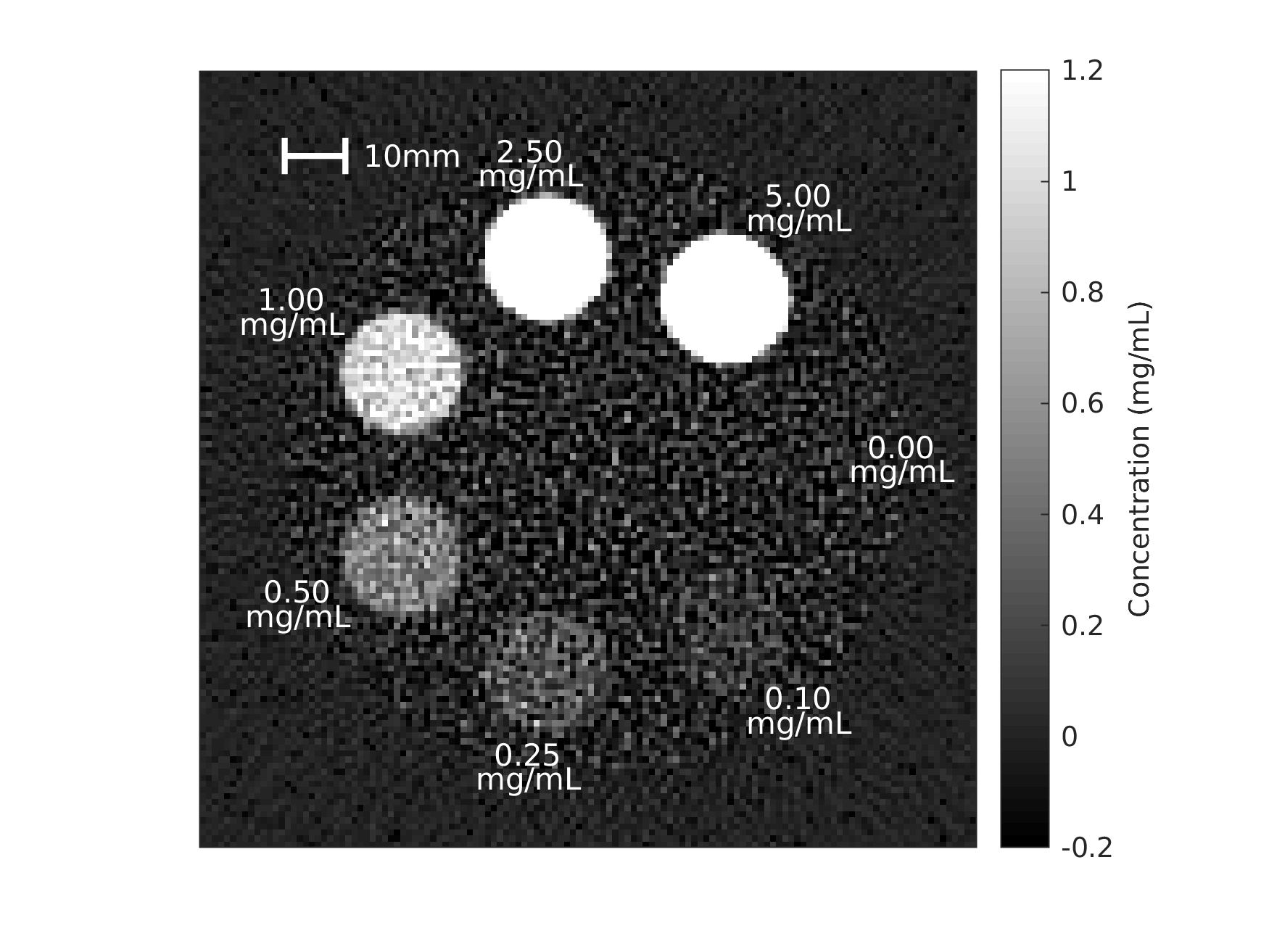}
    \includegraphics[height=0.34\textwidth,trim = {24mm 2mm 12mm 0},clip]{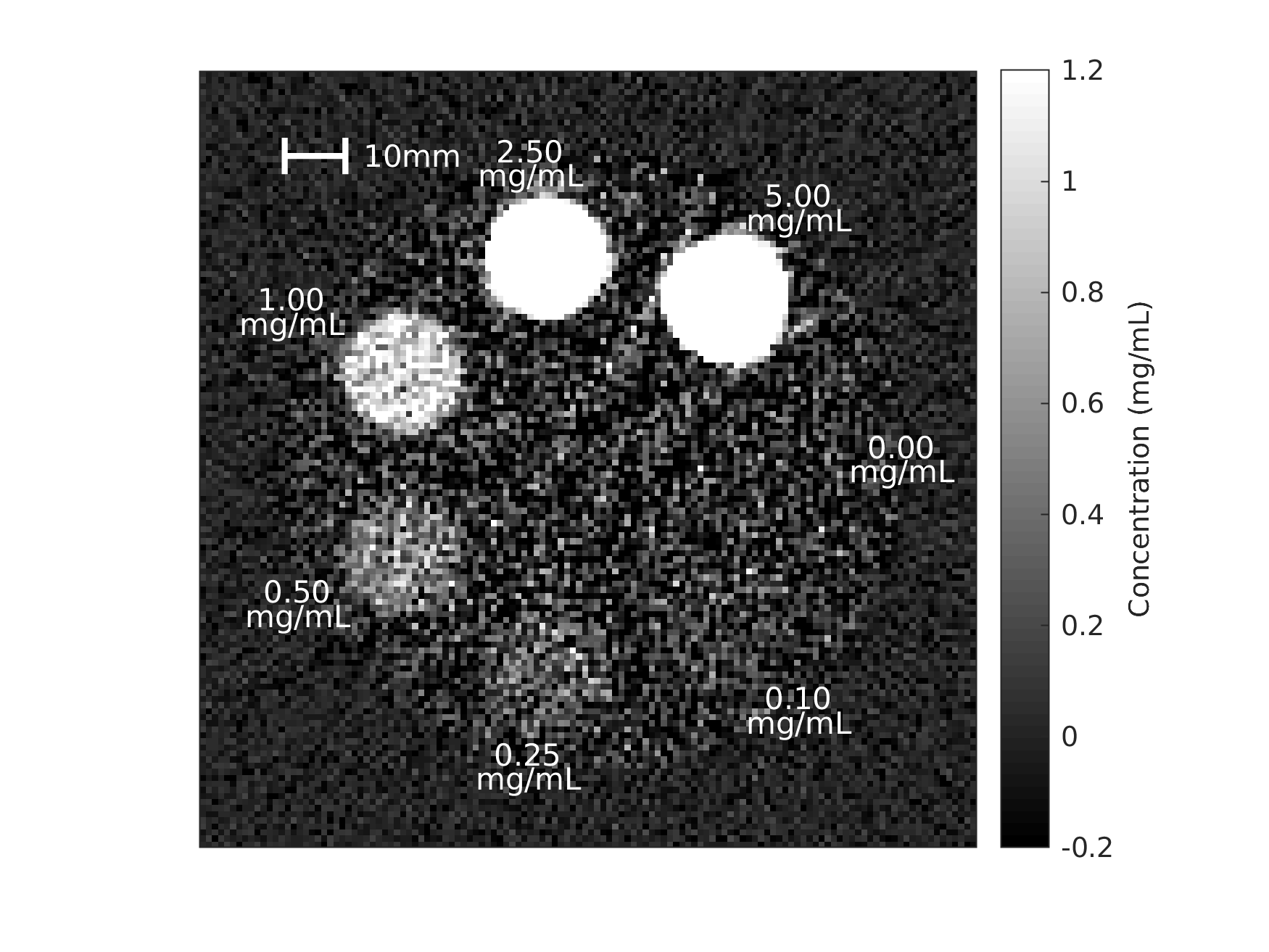}
    \caption{Iodine density maps estimated via MBMD. KVS-DLD-SSF with a wide display window (left)  KVS-DLD-SSF with a narrow display window (center) SSF with a narrow display window (right)}
    \label{fig:phantom_recon}
\end{figure}

\vspace{-2mm}

\FloatBarrier

\section{RESULTS}

We present iodine decomposition results in Figure \ref{fig:phantom_recon}. The left-hand image shows that concentrations between 0.10 and 5.00 mg/mL of iodine can be reconstructed with spectral CT. However, we are specifically interested in the low-concentration inserts. We show an image comparison between the hybrid method, KVS-DLD-SSF (center), and the individual method, SSF (right). For this image comparison we matched bias metrics and compared contrast-to-noise ratio as a performance metric for the inserts. The contrast-to-noise ratio for the 0.25 mg/mL insert is 4.17 for KVS-DLD-SSF and 1.03 for SSF, and the contrast-to-noise ratio for the 0.10 mg/mL insert is 1.67 for KVS-DLD-SSF and 0.38 for SSF.

\begin{figure}[h!]
    \centering
    \vspace{-2mm}
    \includegraphics[width=0.9\textwidth,trim = {10mm 2mm 20mm 0}]{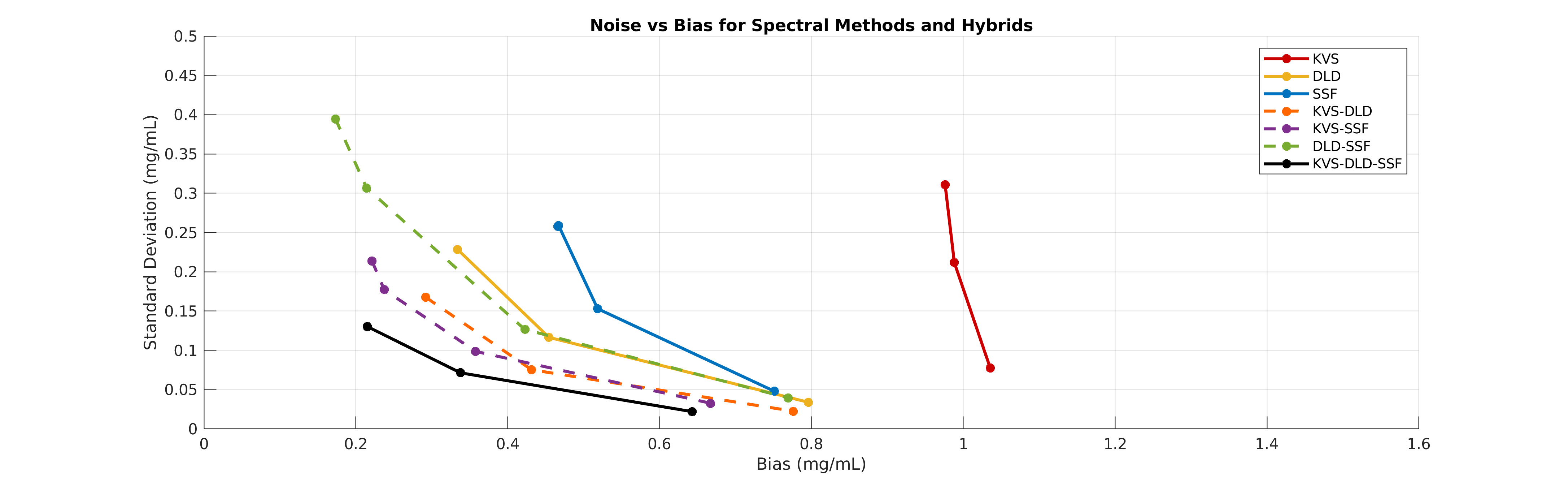}
    \caption{Noise-bias curves for spectral CT methods}
    \label{fig:noise_bias}
\end{figure}



Image quality results across all methods are summarized in the bias-noise plot in Figure \ref{fig:noise_bias}. As expected, greater regularization strength leads to lower noise and higher bias for all methods. We note that individual methods were not optimized - e.g., a different KVS scheme may be more optimal for water-iodine decomposition. The results illustrate that combinations of two strategies have superior performance as compared to the methods acting individually, and that the three-method combination has best overall performance in the study. This suggests that increased spectral diversity has a positive impact on material decomposition performance. 


\section{CONCLUSION}

The search space for optimal hardware designs and configurations is large and we do not make a claim that the settings for individual methods used here are optimal for water-iodine decomposition. Furthermore, what is optimal for individual methods may not be optimal for the hybrid method which combines them. However, we have illustrated that hybrid acquisition methods that achieve greater spectral diversity have a beneficial impact on material decomposition performance. Specifically, for matched bias, we observed higher sensitivity when combining spectral CT methods than any constituent method acting individually. Thus, hybrid acquisition strategies for spectral CT shows promise for the clinically important task of high-sensitivity imaging in contrast-enhanced CT. Future studies will include optimization of individual spectral protocols to further improve low concentration contrast estimates.

\acknowledgments 
This work was supported, in part, by NIH grant R21EB026849.

\bibliography{report}{}
\bibliographystyle{spiebib-abbr} 

\end{document}